\documentstyle[aps,epsfig,multicol]{revtex}

\begin{document}
\draft
\title{Itinerant Ferromagnetism for Mixed Valence Systems}
\author{C. D. Batista,$^1$ J. Bon\v ca,$^2$ and J. E. Gubernatis$^1$}
\address{$^1$Center for Nonlinear Studies and Theoretical Division
Los Alamos National Laboratory, Los Alamos, NM 87545\\
$^2$ Department of Physics, FMF, University of Ljubljana and J.
Stefan Institute, Ljubljana, Slovenia}
\date{\today}
\maketitle
\begin{abstract}

We introduce a novel mechanism for the unusual itinerant
ferromagnetism found in mixed valence systems like
Ce(Rh$_{1-x}$Ru$_x$)$_3$B$_2$ , La$_x$Ce$_{1-x}$Rh$_3$B$_2$ , US,
USe, and UTe.
With it we can provide an explanation for the long-unexplained
large value of $T_c$ ($\sim$ 100$^\circ$K) value and the maximum
in the magnetization below $T_c$ found experimentally.
We also show that this novel
itinerent ferromagnetism can be continuously connected with the
localized case for which the energy scale is much smaller
($J_{RKKY} \sim$ 1$^\circ$K).

\end{abstract}

\begin{multicols}{2}

\columnseprule 0pt

\narrowtext
{\it Introduction.} Understanding the mechanisms for
ferromagnetism usually involves a reconciliation of a localized
electron picture, traced back to Heisenberg \cite{heisenberg}, and
an itinerant electron picture, traced back to Bloch \cite{bloch}.
For many insulating materials, adding itinerant features, such as
indirect exchange, to the localized picture brings satisfactory
agreement with basic experimental features. For many metals,
adding localized features, such as spin waves, to the itinerant
picture has a similar effect. On the other hand it is now
appreciated that for some novel classes of materials, such as the
heavy Fermion and mixed valence materials, doctoring one picture
or the other is often a questionable procedure. The electrons can
display both localized and itinerant properties as these materials
are neither good insulators nor good metals.

In this letter we propose a simple mechanism for ferromagnetism
(FM) in heavy fermion mixed valence materials. The mechanism
relies on several specific energy scales set by the band structure
of these materials even though the electron correlation energy is
the dominant energy scale. The simplicity and generality of the
mechanism however should provide a useful framework for discussing
ferromagnetism in broad classes of heavy fermion mixed valence
materials often found in 4$f$ and 5$f$ materials and their compounds.
In fact we will show that it seems to explain some of the very
unusual long unexplained ferromagnetic properties of such Ce based
compounds as Ce(Rh$_{1-x}$Ru$_x$)$_3$B$_2$ \cite{malik}
and the uranian monochalcogenides \cite{santini}.

The mechanism is based upon a two-band structure. The lower band
near the zone center is dispersive and nearly free of electron
correlation effects. Away from the zone center, this band is very
flat and doubly occupied states experience strong correlation
effects. The upper band is separated from the lower band by a gap
that is smaller than the correlation energy. When the chemical
potential is at the flat band level, the system becomes unstable
to the formation of a ferromagnetic state. There are compounds
satisfying this mixed valence requirement under normal conditions
or  by either adding additional electrons (doping), applying
pressure, or changing the temperature. This ferromagnetic state is
a manifestation of a Hund-like rule among electrons in band states as
opposed to usual case of electrons in localized states (atomic
orbitals).

We will show that the periodic Anderson model (PAM) is a simple
microscopic realization of this picture. Using quantum Monte Carlo
simulations, we will demonstrate that the model admits a
ferromagnetic ground state in the mixed valence regime with
electron occupancies reflecting our physical picture. We then
develop a mean field picture that reproduces the ground state
properties of the PAM and also allows us to compute finite
temperature properties. Under certain conditions we find the
finite temperature mean-field approximation predicts the same highly unusual
behavior as a peak, below $T_c$, in the temperature dependence of
the magnetization \cite{malik,santini} and the deviation of 
the inverse susceptibility from Curie-Weiss behavior \cite{cheche}.

{\it Model Hamiltonian.} The PAM is described by the Hamiltonian
\begin{eqnarray}
  H &=& -t\sum_{\langle i,j \rangle,\sigma} (d_{i\sigma}^\dagger
  d^{}_{j\sigma}+d_{j\sigma}^\dagger d^{}_{i\sigma})
  +V\sum_{i,\sigma} (d_{i\sigma}^\dagger
  f^{}_{i\sigma}+f_{i\sigma}^\dagger d^{}_{i\sigma}) \nonumber \\ & &
  \quad\quad +\epsilon_f\sum_{i,\sigma}n_{i\sigma}^f +\frac{U}{2}
  \sum_{i,\sigma}n_{i\sigma}^fn_{i\bar {\sigma}}^f
\label{eq:pam}
\end{eqnarray}
where $d_{i\sigma}^\dagger$ and $f_{i\sigma}^\dagger$ create an
electron with spin $\sigma$ in $d$ and $f$ orbitals at lattice
site $i$ and $n^f_{i\sigma}=f^{\dagger}_{i\sigma}f^{}_{i\sigma}$
is the number operator for the $f$-electrons of spin ${\sigma}$ at
site $i$. The hopping amplitude $t$
between $d$-orbitals is only to nearest-neighbor sites. The
hopping amplitude $V$ hybridizes different orbitals on the same
site.

When $U=0$, the resulting non-interacting Hamiltonian $H_0$ has
two dispersive bands:
\begin{equation}
 E_\sigma^\pm({\bf k})=\frac{1}{2} \Biggl[
   e_{\bf k}+\epsilon_f \pm \sqrt{(e_{\bf k}
                                            -\epsilon_f)^2+4V^2}
   \Biggr]
\end{equation}
separated by an hybridization gap. 
For a chain, the band structure is illustrated in Fig.~1a.
Irrespective of spatial dimension, each band has regions where the
${\bf k}$-states are dominantly of $d$ or $f$-character. A mixed valence
condition arises when the Fermi energy $E_F$ sits in a crossover
region which is around the energy $\epsilon_f$. Three energies are
important: the hybridization gap $\Delta$, the $f$-state
dispersion $\delta_f=\partial E^{-}/\partial k|_{k_F}=\hbar v_F$ ($v_F$
is the Fermi velocity), 
and the Coulomb (correlation) energy $U$.
$\delta_f$  is measuring  the band dispersion close to the Fermi level. In the 
interacting problem we will
assume that $U>\delta_f,\Delta$. As we shall now discuss, the
competition is between $\delta_f$ and $\Delta$ which will be a
competition between a ferromagnetic and paramagnetic state.

{\it Mechanism for Ferromagnetism.} In Fig.~1a-c we illustrate the
physical mechanism for the ferromagnetic state which emerges when
the electron filling $\rho$ is arround 1/4 
(one electron per lattice site) 
and the energy $\epsilon_f$ of the $f$-orbitals is
close to $E_F$. The latter condition defines the intermediate
valence regime.

To explain the mechanism, we first consider the non-interacting
case (Fig.~1a). States in the lower band with mainly $f$-character
have a small dispersion $\delta_f$ because of the absence of
direct hopping between the $f$-orbitals. In the interacting case,
because the Coulomb interaction only affects the electrons in the
$f$-orbitals, the electrons which were doubly occupying the states
in the lower band with mainly $d$-character are practically
unaffected. On the other hand, the electrons in the many states
which are close to the Fermi level and have mainly an
$f$-character will be strongly affected. These electrons spread to
higher unoccupied ${\bf k}$-states in the $f$-part of the band and
polarize by an in-band Hund's rule (Fig.~1b). To see the analogy
to Hund's rule more clearly, we consider the limiting case,
represented in Fig.~1c, where $\delta_f=0$ and $\Delta \neq 0$,
reducing Fig.~1b to a two level system in momentum space with each
level being strongly degenerate (we only plot the states with predominant $f$
character). If we add electrons to this two levels system 
it is easy to show that the ferromagnetic solution has the lowest energy and is
therefore the ground state. By polarizing, the spatial part of the 
wave function becomes antisymmetric and there is no double occupancy 
in the real space. In this way, the Coulomb repulsion is reduced 
to zero. The kinetic energy has also the lowest possible value if the electrons
are occupying the lower energy (degenerated) levels (see Fig.~1c).

We now describe the conditions for the stability of the 
ferromagnetic state, and we will see how this 
mechanism departs from the one of Stoner \cite{stoner}.
To this end we state the condition $U \gg \Delta$. From Fig.~1b
it is clear that the cost of the spreading of the ferromagnetic 
solution is an increase in the kinetic energy proportional to $\delta_f$.
On the other hand, if we build up a nonmagnetic solution only
using the states of the lower band, there will be a minimum delocalization
for each particle due to the fact that a finite set of $f$-character 
states is in the upper band (see Fig.~1). This can be seen by 
constructing an $f$ Wannier function with the lower band states.
This Wannier function will have a delocalization lenght which depends 
on the ${\bf k}$ wave vector where the two bands are crossing.  
Therefore, by localizing in real space to avoid the double occupancy, 
the electrons will have a finite
probability of occupying the $f$-states in the upper band. The
energy cost per electron of occupancy is proportional to $\Delta$.
If the hybridization gap is much larger than the dispersion of the
states, that is, $\Delta \gg \delta_f$, the ferromagnetic state lies
lower in energy than the nonmagnetic state. The energy of the
excited nonmagnetic state is proportional to $\Delta$. This can
be easily seen from the limiting (and non-realistic) case $\delta_f=0$ 
ploted in Fig.~1c, for which the analogy with the atomic shells and
the Hund's rule is evident.

\begin{figure}[tbp] 
\begin{center}
\vspace{-1.0cm}
\epsfig{file=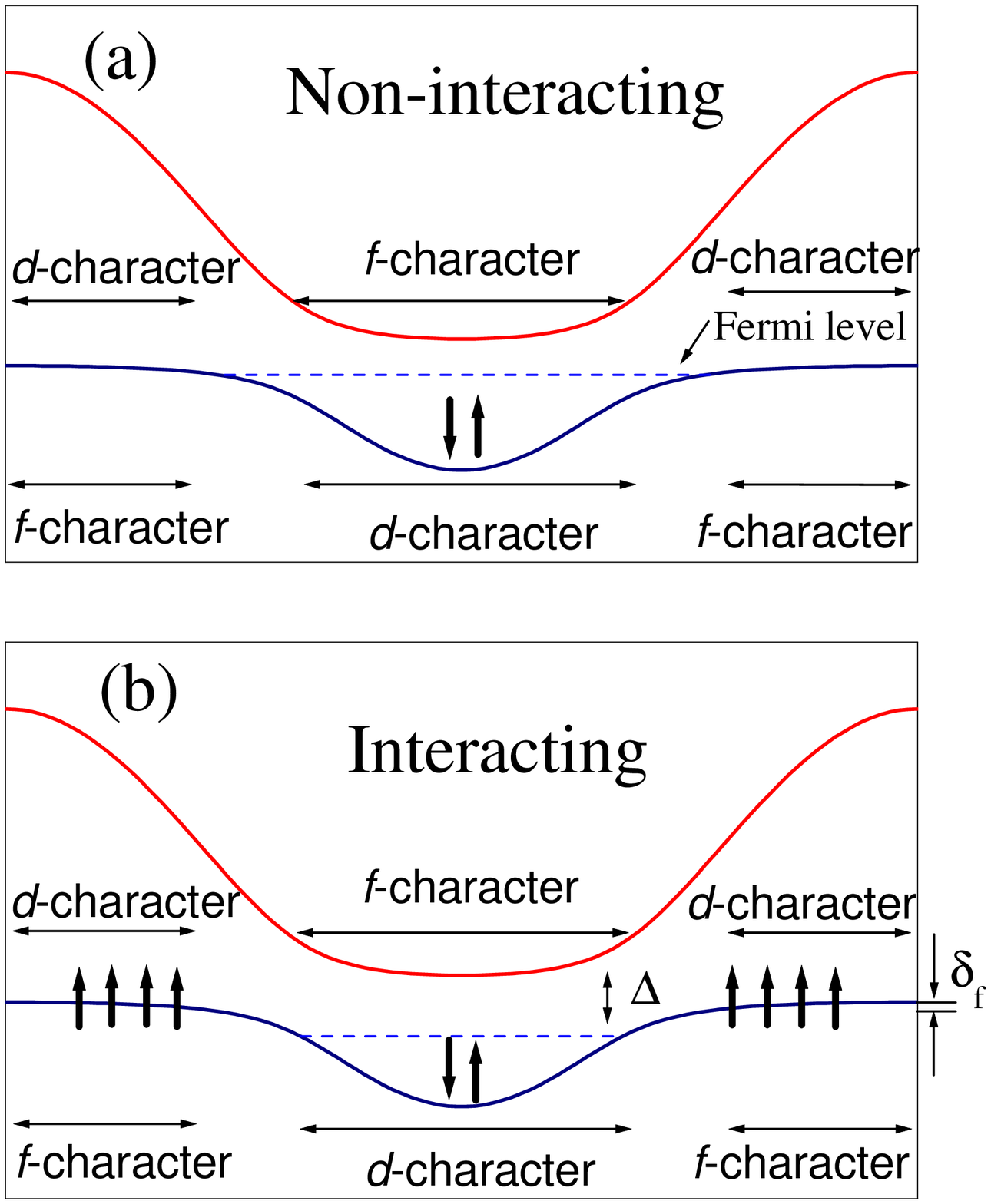,width=70mm,angle=-0}
\end{center}
\end{figure}
\begin{figure}[tbp]
\begin{center}
\vspace{-3.0cm}
\epsfig{file=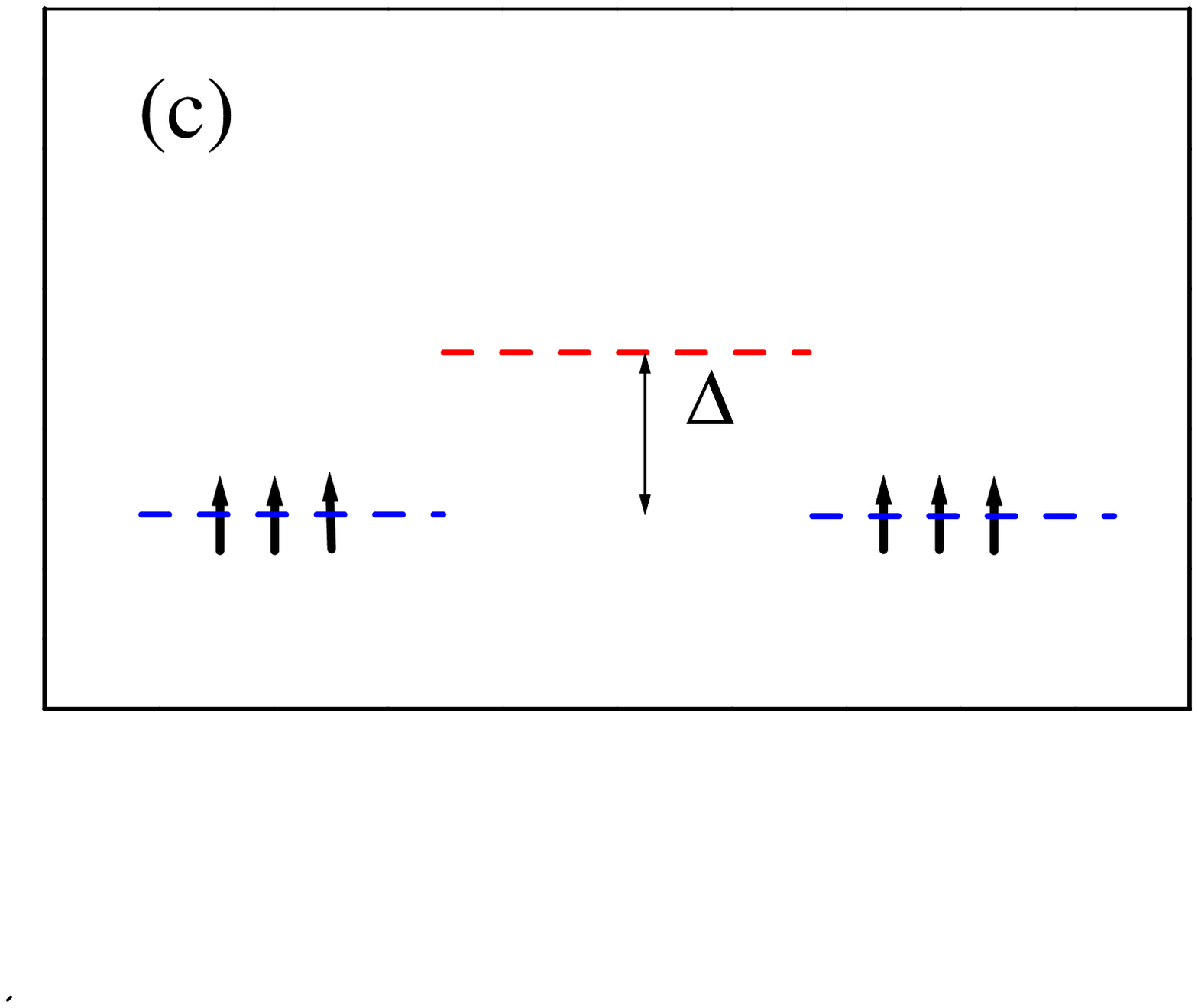,width=68mm,angle=-0}
\vspace{-5.1cm}
\end{center}
\caption{Illustration of the ferromagnetic mechanism.}
\label{fig1}
\end{figure}

It is important to emphasize that the hybridization gap $\Delta$
and $f$-state dispersion $\delta_f$ are the two basic ingredients
of our mechanism: $\Delta$ only appears if there is more than one
band and $\delta_f$ is small only if the hybridization is weak. If
$U>\Delta,\delta_f$, then $\Delta$ is the energy scale of the
paramagnetic state, while $\delta_f$ is the energy scale of the
ferromagnetic solution. We can estimate the magnitude of these
scales for the PAM: for $V \lesssim t/2$, we get $\delta_f \sim
\partial e_k/\partial k|_{k_F} V^2/(e_{k_F}-\epsilon_f)^2$ 
and $\Delta \sim V^2/(2dt-\epsilon_f)+
\sqrt{(2dt+\epsilon_f)^2+4V^2}-2dt$. Therefore, if $\epsilon_f$ is
close to the bottom of the lower band ($\epsilon_f+2dt \sim t$)
and the Fermi level is close to the top ($E_F \sim \epsilon_f-V^2/(2dt-\epsilon_f)$),
$\Delta$ is considerably larger than $\delta_f$ and the ground
state is ferromagnetic. We can see that the ferromagnetic
solution is stable for comfortably realistic values of the
parameters. It is clear from this analysis the particle density $\rho$
for a ferromagnetic solution must be close to quarter filling.

{\it Quantum Monte Carlo Method.} Using the constrained-path Monte
Carlo method (CPMC), we computed the ground-state properties of
the PAM on a square lattice. The CPMC method projects the ground
state from an initial state $|\psi_T\rangle$ by converting the
iterative procedure
\begin{equation}\label{eq:cpmc}
  |\psi_{i+1}\rangle =e^{-\tau H}|\psi_{i}\rangle
\end{equation}
into a branched random walk. The details of the method are
described elsewhere \cite{zhang} as are our prior uses of it on
the square PAM \cite{bonca,batista}.

The defining characteristic of the CPMC method is its elimination
of the fermion sign problem by excluding random walkers
$|\phi\rangle$ that violate $\langle \psi_T|\phi\rangle>0$. If
$|\psi_T\rangle$ were the exact ground state $|\psi_0\rangle$,
this constraint would generate an exact elimination of the sign
problem, and hence an exact solution. Extensive benchmarking
indicates the CPMC method provides accurate estimates of the
energy and various correlation functions.

We prepared the trial state in specific values of the total spin
and z-component of spin \cite{batista}. Because the
Hamiltonian(\ref{eq:pam}) conserves these quantum numbers, the
iterative process (\ref{eq:cpmc}) produces a ground state with the
same $S$ and $S_z$. In the ferromagnetic regime, when the number of
lattice sites $N$ was greater than the number of electrons $N_e$,
the resulting energy as a function of $S$ almost always showed a
minimum at a value of $S$ satisfying
$0<S<\frac{1}{2}(2N-N_e)$. Hence we typically found a partially
polarized ferromagnetic ground state.

For square lattices we also computed dependence of the static spin structure
factor and the electron occupancies both on position and wave
number space \cite{bonca}. Most notable for the present work is
the wave vector dependence of the different spin components of the
upper and lower band electron occupancies shown in Fig.~2. It has
the same quantitative features as Fig.~1b.

{\it Mean Field Theory.} The quantum Monte Carlo simulations are
limited to chains and square lattices of relatively small sizes at
zero temperature. However, we have found that their predictions
are well described by a simple spin-polarized mean-field
approximation. We can easily extend this approximation to larger
systems sizes in higher dimensions and at finite temperatures. We
can use large lattices in three-dimensions at finite temperatures
to compare with the results of experiments.


The mean field Hamiltonian $H_{MF}$ is
\begin{equation}
  H_{MF} = H_{0} + \frac{U}{2}
\sum_{i,\sigma} (\langle n_{i\sigma}^f \rangle n_{i\bar
{\sigma}}^f + n_{i\sigma}^f \langle n_{i\bar {\sigma}}^f \rangle-
\langle n_{i\sigma}^f \rangle  \langle n_{i\bar {\sigma}}^f
\rangle) \label{eq:mf}
\end{equation}
$\langle n_{i \uparrow}^f \rangle$ and $\langle n_{i \downarrow}^f
\rangle$ are the determined in a self-consistent way within the
mean-field theory. The translational symmetry of $H$ implies that
we can always find a translationally invariant ground state. This
means that $\langle n_{i\uparrow}^f \rangle$ and $\langle n_{i
\downarrow}^f \rangle$ will not depend on the site index $i$.
Therefore there are two variational parameters to be determined
from the mean-field equations: $\langle n_{\uparrow}^f \rangle$
and $\langle n_{\downarrow}^f \rangle$.

Even though the mean field approximation can give a good
description of the FM ground state, it is well known that such
approximations overestimate the energy of the paramagnetic phase
because they improperly estimate the real space correlations that
are very important for the paramagnetic solution. Therefore it is
crucial to check that the ground state is the FM one by using a
more accurate method to evaluate the energy of the PM state. To
this end we calculated the ground state of the PAM with $6 \times
6$ unit cells using the CPMC technique. We
found excellent agreement with the energy and as seen in Fig.~2
the electron occupancies.

\begin{figure}[tbp]
\begin{center}
\vspace{-1.2cm}
\epsfig{file=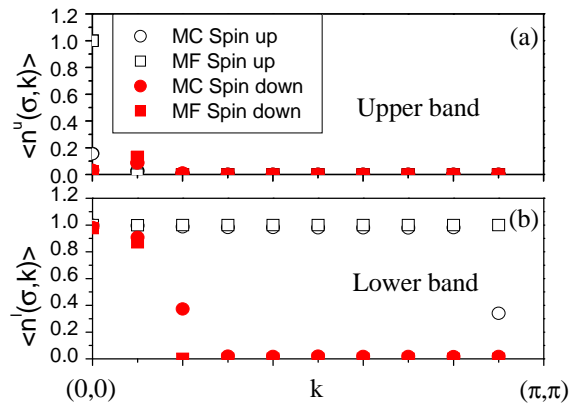,width=80mm,angle=-90}
\vspace{-2.3cm}
\end{center}
\caption{Comparison between Mean Field and CPMC values of the 
occupation numbers of the non-interacting band states
for each spin polarization in a $6\times6$ cluster
($\rho=7/24$,$V=0.5t$, $U=t$, and $\epsilon_f=-3t$). On the 
$x$-axis we only include the nonequivalent wave vectors ordered
according to increasing non-interacting energies.}
\label{fig2}
\end{figure}

{\it Consequences.} Bolstered by this agreement we used the
mean-field approximation to calculate various thermodynamic
properties of the ferromagnetic phase for the three-dimensioanl
cubic PAM. In Fig.~3a we show the temperature dependence of the
magnetization for different values of $\epsilon_f$. From the
arguments given above it is clear that the magnetization comes
from the electrons which are occupying the states with
$f$-character. By increasing $\epsilon_f$ we are reducing the
number of $f$-electrons and therefore the magnitude of the zero
temperature magnetization. Some critical value of $\epsilon_f$
exists where the chemical potential starts departing from
$\epsilon_f$ and a new energy scale $\epsilon_f-E_F$ emerges in
the system. This new energy scale is reflected in the appearance
of a magnetization peak. (See the $\epsilon_f=-t$ case in
Fig.~3a.) To understand the last statement one first has to
realize that the zero temperature magnetization is small due to
the reduction of the number of $f$-electrons. When the temperature
is of the order of $\epsilon_f-E_F$, electrons are promoted
electrons from the double occupied band states to the
$f$-character states which have a large entropy (large density of
states). The $f$-electrons will be polarized due to the energy
considerations above explained. In this way we can explain the
origin of the magnetization peak below $T_c$. It is important to
note that the source for the large entropy is associated with
charge and not with spin degrees of freedom. This fact explains
why a state with larger magnetization has a higher entropy. The
magnetization curves shown in Fig. 3a are in good qualitative
agreement with the magnetization versus temperature data measured
in Ce(Rh$_{1-x}$Ru$_{x}$)$_3$B$_2$ for different values of $x$
\cite{malik}.


\begin{figure}[tbp]
\begin{center}
\vspace{-1.2cm}
\epsfig{file=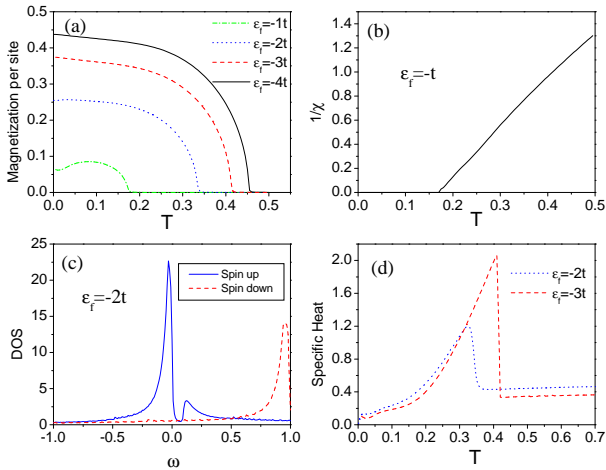,width=90mm,angle=-90}
\vspace{-2.5cm}
\end{center}
\caption{Mean field results ($\rho=1/4$,$V=0.5t$, $U=2t$) for (a) magnetization,
(b) inverse susceptibility,
(c) density of states and (d) specific heat.}
\label{fig3}
\end{figure}

Another interesting aspect of this ferromagnetism is the deviation
from linearity (Curie-Weiss behavior)
for the inverse magnetic susceptibility above
$T_c$ over a large temperature range (see Fig. 3b) \cite{cheche}. 
The behavior contrasts that predicted by theories based on 
localized magnetic moments.

From above  it is also clear that the change in entropy from the
magnetic to the paramagnetic phase depends on the number for $f$
electrons (spin degrees of freedom). This dependence is seen in
Fig.~3d where we plot the calculated specific heat for different
values of $\epsilon_f$.

A characteristic of this ferromagnetic state that has consequences
for the photoemission experiments is the implication that the
quasi-particle dispersion should be close to that of the electrons
in  the non-interacting case. We recall that the ferromagnetic
state described in Fig. 1 is very similar to a non-interacting
polarized state. As in the Stoner mechanism, the main role of the
Coulomb repulsion is to polarize states with well defined
momentum. The hybridization gap of the noninteracting solution is
most likely replaced by a pseudogap (see Fig.~3c). This pseudogap should be seen
in the optical conductivity measurements.

{\it Conclusions.} Our quantum Monte Carlo simulations and
mean-field calculations clearly show the existence of an itinerant
ferromagnetic phase in mixed valence materials that is supported
by the PAM. This new phase describes qualitatively  many experimental
features such as an unusually large value for $T_c$($\sim$
100$^\circ$K) and the maximum in the magnetization below $T_c$
\cite{shaheen} that are found in such ferromagnetic compounds as
Ce(Rh$_{1-x}$Ru$_x$)$_3$B$_2$ \cite{malik},
Ce(Rh$_{1-x}$Os$_x$)$_3$B$_2$ \cite{malik}, and
La$_x$Ce$_{1-x}$Rh$_3$B$_2$ \cite{shaheen}.

Many previous theories of 4$f$ and 5$f$ electron materials treated
these materials as systems of localized moments in the
$f$-orbitals. These theories are thus unable to describe an
itinerant ferromagnetic phase and experimental consequences
\cite{santini} like the peak in the magnetization below $T_c$
observed in some of these systems \cite{malik,cheche}, the large
value of $T_c$, the deviation of $\chi^{-1}(T)$ from the
Curie-Weiss law above $T_c$ in the uranium monochalcogenides
\cite{cheche}, and the mixed valence behavior of these compounds
\cite{shaheen,kanter,erbudak}.

The physical picture just presented, when combined with our
previous results \cite{batista}, allows a reconciliation of the
localized and delocalized ferromagnetism pictures painted by
Heisenberg and Bloch seventy years ago in the sense that it is
possible to go continuously from the mixed valence FM state where
the $f$-electrons are delocalized to a FM state where there is one
localized electron in each $f$-orbital \cite{batista}. In our
picture one can do this by decreasing $\epsilon_f$ from the Fermi
level to a value near or below the bottom of the valence band. We
note that the energy scale of a localized ferromagnetic state is
that of the RKKY interaction, which according to de Gennes's rule
\cite{degennes} is of the order of 1$^\circ$K, while the scale for
the itinerant case introduced here is that of the hybridization
gap $\Delta$, which is of the order of 100$^\circ$K for heavy
fermion compounds \cite{note}. In going from the delocalized (mixed valence)
to the localized regime we thus expect a strong reduction of $T_c$
accompanied by an increase of the zero temperature magnetization.
(The $T=0$ magnetization is proportional to the $f$ occupation
number.) This expectation is consistent with the observed behavior
of the magnetization in La$_x$Ce$_{1-x}$Rh$_3$B$_2$ as function of
$x$ \cite{shaheen}.

{\it Acknowledgements.} This work was sponsored by the US DOE
under contract W-7405-ENG-36. We acknowledge useful discussions
with A. J. Arko, B. Brandow, M. F. Hundley, J. J. Joyce,   
J. M. Lawrence, S. Trugman, and J. L. Smith. In particular, 
we thank J. M. Lawrence 
for pointing out the experimental work on the Ce compounds.
J. B. 
acknowledges the support Slovene Ministry of Education, Science and
Sport, Los Alamos National Laboratory and FERLIN.


\end{multicols}


\begin{thebibliography}{99}
\bibitem{note}
This change in the energy scale of both ferromagnetic states
is obtained with Monte Carlo by taking the energy difference between 
the PM and the FM solutions at $T=0$. The mean field theory
is not appropiate to describe this transition because it overestimates
the PM energy.
\end{thebibliography}
\end{document}